\documentclass[twoside,twocolumn,9pt]{article}
\usepackage{extsizes}
\usepackage[super,sort&compress,comma]{natbib} 
\usepackage[version=3]{mhchem}
\usepackage[left=1.5cm, right=1.5cm, top=1.785cm, bottom=2.0cm]{geometry}
\usepackage{balance}
\usepackage{mathptmx}
\usepackage{sectsty}
\usepackage{graphicx} 
\usepackage{lastpage}
\usepackage[format=plain,justification=justified,singlelinecheck=false,font={stretch=1.125,small,sf},labelfont=bf,labelsep=space]{caption}
\usepackage{float}
\usepackage{fancyhdr}
\usepackage{fnpos}
\usepackage[english]{babel}
\addto{\captionsenglish}{%
  
}
\usepackage{array}
\usepackage{droidsans}
\usepackage{charter}
\usepackage[T1]{fontenc}
\usepackage[usenames,dvipsnames]{xcolor}
\usepackage{setspace}
\usepackage[compact]{titlesec}
\usepackage{hyperref}
\usepackage{epstopdf}%This line makes .eps figures into .pdf - please comment out if not required.

\definecolor{cream}{RGB}{222,217,201}

\begin{document}

\pagestyle{fancy}
\thispagestyle{plain}
\fancypagestyle{plain}{
%%%HEADER%%%
\renewcommand{\headrulewidth}{0pt}
}
%%%END OF HEADER%%%

%%%PAGE SETUP - Please do not change any commands within this section%%%
\makeFNbottom
\makeatletter
\renewcommand\LARGE{\@setfontsize\LARGE{15pt}{17}}
\renewcommand\Large{\@setfontsize\Large{12pt}{14}}
\renewcommand\large{\@setfontsize\large{10pt}{12}}
\renewcommand\footnotesize{\@setfontsize\footnotesize{7pt}{10}}
\makeatother

\renewcommand{\thefootnote}{\fnsymbol{footnote}}
\renewcommand\footnoterule{\vspace*{1pt}% 
\color{cream}\hrule width 3.5in height 0.4pt \color{black}\vspace*{5pt}} 
\setcounter{secnumdepth}{5}

\makeatletter 
\renewcommand\@biblabel[1]{#1}            
\renewcommand\@makefntext[1]% 
{\noindent\makebox[0pt][r]{\@thefnmark\,}#1}
\makeatother 
\renewcommand{\figurename}{\small{Fig.}~}
\sectionfont{\sffamily\Large}
\subsectionfont{\normalsize}
\subsubsectionfont{\bf}
\setstretch{1.125} %In particular, please do not alter this line.
\setlength{\skip\footins}{0.8cm}
\setlength{\footnotesep}{0.25cm}
\setlength{\jot}{10pt}
\titlespacing*{\section}{0pt}{4pt}{4pt}
\titlespacing*{\subsection}{0pt}{15pt}{1pt}
%%%END OF PAGE SETUP%%%

%%%FOOTER%%%
\fancyfoot{}
\fancyfoot[LO,RE]{\vspace{-7.1pt}\includegraphics[height=9pt]{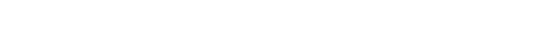}}
\fancyfoot[CO]{\vspace{-7.1pt}\hspace{13.2cm}\includegraphics{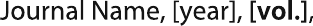}}
\fancyfoot[CE]{\vspace{-7.2pt}\hspace{-14.2cm}\includegraphics{head_foot/RF}}
\fancyfoot[RO]{\footnotesize{\sffamily{1--\pageref{LastPage} ~\textbar  \hspace{2pt}\thepage}}}
\fancyfoot[LE]{\footnotesize{\sffamily{\thepage~\textbar\hspace{3.45cm} 1--\pageref{LastPage}}}}
\fancyhead{}
\renewcommand{\headrulewidth}{0pt} 
\renewcommand{\footrulewidth}{0pt}
\setlength{\arrayrulewidth}{1pt}
\setlength{\columnsep}{6.5mm}
\setlength\bibsep{1pt}
%%%END OF FOOTER%%%

%%%FIGURE SETUP - please do not change any commands within this section%%%
\makeatletter 
\newlength{\figrulesep} 
\setlength{\figrulesep}{0.5\textfloatsep} 

\newcommand{\topfigrule}{\vspace*{-1pt}% 
\noindent{\color{cream}\rule[-\figrulesep]{\columnwidth}{1.5pt}} }

\newcommand{\botfigrule}{\vspace*{-2pt}% 
\noindent{\color{cream}\rule[\figrulesep]{\columnwidth}{1.5pt}} }

\newcommand{\dblfigrule}{\vspace*{-1pt}% 
\noindent{\color{cream}\rule[-\figrulesep]{\textwidth}{1.5pt}} }

\makeatother
%%%END OF FIGURE SETUP%%%

%%%TITLE, AUTHORS AND ABSTRACT%%%
\twocolumn[
  \begin{@twocolumnfalse}
{\includegraphics[height=30pt]{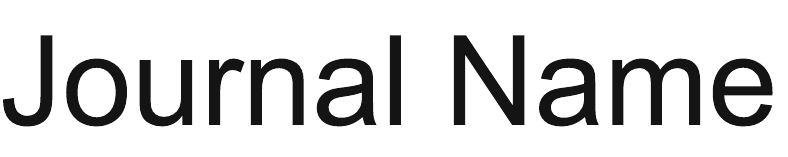}\hfill\raisebox{0pt}[0pt][0pt]{\includegraphics[height=55pt]{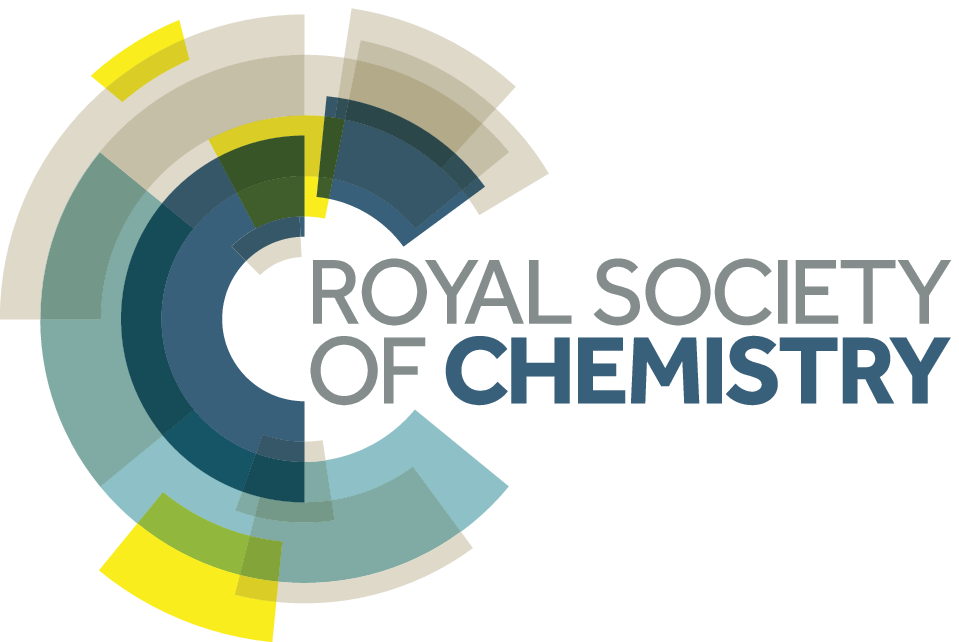}}\\[1ex]
\includegraphics[width=18.5cm]{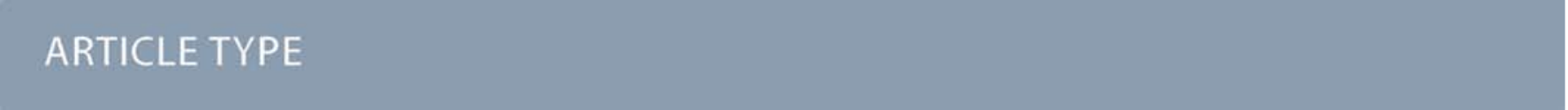}}\par
\vspace{1em}
\sffamily
\begin{tabular}{m{4.5cm} p{13.5cm} }

\includegraphics{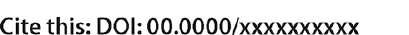} & \noindent\LARGE{\textbf{Structural evolution of binary oxide nanolaminates with annealing and its impact on room-temperature internal friction}} \\%Article title goes here instead of the text "This is the title"
\vspace{0.3cm} & \vspace{0.3cm} \\

 & \noindent\large{Le Yang,$^{\ast}$\textit{$^{a}$} Mariana Fazio,\textit{$^{b}$} and Gabriele Vajente,\textit{$^{c}$}Alena Ananyeva,\textit{$^{c}$}GariLynn Billingsley, \textit{$^{d}$}Ashot Markosyan,\textit{$^{d}$}Riccardo Bassiri,\textit{$^{d}$} Martin M. Fejer,\textit{$^{d}$} Carmen S. Menoni\textit{$^{b\ddag}$}} \\%Author names go here instead of "Full name", etc.

\includegraphics{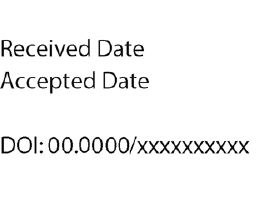} & \noindent\normalsize{Internal friction in oxide thin films imposes a critical limitation to the sensitivity and stability of ultra-high finesse optical cavities for gravitational wave detectors. Strategies like doping or creating nanolaminates are sought to introduce structural modifications that reduce internal friction. This work describes an investigation of the morphological changes SiO$_2$/Ta$_2$O$_5$ and TiO$_2$/Ta$_2$O$_5$ nanolaminates undergo with annealing and their impact on room temperature internal friction. It is demonstrated that thermal treatment results in a reduction of internal friction in both nanolaminates, but through different pathways. In the SiO$_2$/Ta$_2$O$_5$ nanolaminate, which layers remain intact after annealing, the total reduction in internal friction follows the reduction in the composing SiO$_2$ and Ta$_2$O$_5$ layers. Instead, interdiffusion initiated by annealing at the interface of the TiO$_2$/Ta$_2$O$_5$ nanolaminate and the formation of a  mixed phase dictate a more significant reduction in internal friction to $\sim$ 2.6 $\times$ 10$^{-4}$, a value lower than any other Ta$_2$O$_5$ mixture coating with similar cation concentration.} \\%The abstrast goes here instead of the text "The abstract should be..."

\end{tabular}

 \end{@twocolumnfalse} \vspace{0.6cm}

  ]
%%%END OF TITLE, AUTHORS AND ABSTRACT%%%

%%%FONT SETUP - please do not change any commands within this section
\renewcommand*\rmdefault{bch}\normalfont\upshape
\rmfamily
\section*{}
\vspace{-1cm}

%%%FOOTNOTES%%%

\footnotetext{\textit{$^{a}$~Department of Chemistry, Colorado State University, Fort Collins, CO 80523, USA; E-mail: yangle@colostate.edu}}
\footnotetext{\textit{$^{b}$~Department of Electrical and Computer Engineering, Colorado State University, Fort Collins, CO 80523, USA}}
\footnotetext{\textit{$^{c}$~LIGO Laboratory, California Institute of Technology, Pasadena, CA 91125, USA}}
\footnotetext{\textit{$^{d}$~Edward L. Ginzton Laboratory, Stanford University, Stanford, CA 94305, USA}}

%Please use \dag to cite the ESI in the main text of the article.
%If you article does not have ESI please remove the the \dag symbol from the title and the footnotetext below.
\footnotetext{\dag~Electronic Supplementary Information (ESI) available. See DOI: 00.0000/00000000.}
%additional addresses can be cited as above using the lower-case letters, c, d, e... If all authors are from the same address, no letter is required

\footnotetext{\ddag~Carmen.Menoni@colostate.edu}

%%%END OF FOOTNOTES%%%

%%%MAIN TEXT%%%%

\section{Introduction}
Nanolaminates are representative engineered materials that consist of stacks of nanometer-thick layers of two or more dissimilar materials.  These two-dimensional composite thin films are extensively used in semiconductor devices \cite{tsipas2008germanium, wilk2000hafnium, boscherini2011atomic}, electrochemical storage devices \cite{huo2020two}, and optical coatings \cite{sahoo1998mgo,karvonen2013enhancement}. Their unique properties arise from modifications to the environment of atoms at nanometer to subnanometer scale. For example, in nanolaminates of Al$_2$O$_3$/ZnO \cite{li2017quantum} an exponential increase in emission intensity was achieved due to quantum confinement when the ZnO sublayer thickness was reduced to less than 2 nm. The large interfacial area between layers also plays a critical role in modifying the structural properties of nanolaminates. In HfO$_2$/Al$_2$O$_3$ \cite{wang2011structure} and HfO$_2$/La$_2$O$_3$ \cite{wang2009atomic} nanolaminates it is found that the amorphous state of HfO$_2$ can be stabilized up to 800 $^{\circ}$C annealing, an effect created by the increased contribution from surface enthalpy to the total film energy. At the interface, when chemical reactions occur, advantageous functional properties develop, as is the formation of robust Al-O-Ti and Al-O-Zr bonds that inhibit water incorporation in Al$_2$O$_3$/TiO$_2$ \cite{kim2014al2o3,kim2016optimization} and Al$_2$O$_3$/ZrO$_2$ nanolaminates \cite{meyer2009al2o3}. The improved water anticorrosion is ascribed to the formation of a ternary phase that is thermodynamically more stable than separate phases of the binary oxides \cite{zhao2002miscibility,meyer2010origin}.

The ability to alter structural properties in nanolaminates has also stimulated great interest for their use in highly reflective mirror coatings of ultrastable optical cavities. High finesse cavities are essential components for precision measurements of time and space, such  as atomic  clocks  and  gravitational  wave detectors \cite{cole2013tenfold, morell2016high}. For these applications, reducing thermally driven fluctuation is critical to stabilizing the cavity length and thus improving the system's sensitivity. For example, a factor of four reduction in internal friction (Q$^{-1}$) of the mirror coating in the vicinity of 100 Hz would expand the ability of the Advanced LIGO to detect astrophysical events beyond its present limit, 120 megaparsecs \cite{miller2015prospects,vajente2019precision, abbott2018prospects}. Nanolaminates are envisioned to replace the high index layers in the current mirror coatings. The structural modifications associated with the nanolaminate architecture are expected to alter medium range order of the atomic structure and hence the sources of internal friction, which are conceptualized as two-level systems (TLSs) \cite{billman2017origin}. Experiments on TiO$_2$/SiO$_2$ nanolaminates have shown pronounced suppression of internal friction at cryogenic temperatures \cite{kuo2019low}. This behavior was attributed to the interruption of long scale building blocks associated with TLSs by constraining the layer thickness. In contrast, a direct observation and an unambiguous identification of the structural features that relate to the reduction in room temperature internal friction is still lacking. Identifying and understanding the key structural evolution that governs the internal friction behavior will accelerate the search for coating materials for the next generation gravitational wave detectors.

Herein, we describe the behavior of room temperature internal friction of SiO$_2$/Ta$_2$O$_5$ and TiO$_2$/Ta$_2$O$_5$ nanolaminates with the goal to identify the origin of internal friction reduction in these amorphous oxide thin films. It is shown that the morphological and structural evolution of the SiO$_2$/Ta$_2$O$_5$ nanolaminate upon annealing is significantly different from that of the TiO$_2$/Ta$_2$O$_5$ nanolaminate. The interfaces between layers remain sharp up to an annealing temperature of 650$^{\circ}$C in the SiO$_2$/Ta$_2$O$_5$ nanolaminate. Instead, interdiffusion in the TiO$_2$/Ta$_2$O$_5$ nanolaminate is revealed by high resolution transmission electron microscopy. In both cases, annealing lowers internal friction compared to a single Ta$_2$O$_5$ layer, but from different origin. The reduction in internal friction of the SiO$_2$/Ta$_2$O$_5$ nanolaminate results from the combined reduction in the individual Ta$_2$O$_5$ and SiO$_2$ nanolayers. Instead, interdiffusion and formation of a stable ternary phase upon annealing is responsible for the reduction of internal friction in the TiO$_2$/Ta$_2$O$_5$ nanolaminate.

\section{Experimental}

Nanolaminates of SiO$_2$/Ta$_2$O$_5$ and TiO$_2$/Ta$_2$O$_5$ were prepared by reactive ion beam sputtering using a biased target deposition system \cite{west2008growth}. A high purity metal target is sputtered by Ar ions when a negative bias of 800 V is applied to the target. Ultra-high purity oxygen gas is introduced to the chamber near the substrate surface to grow oxides. Optimization of the oxygen flow was carried out to ensure stoichiometry of the oxides. Deposition rates of Ta$_2$O$_5$, TiO$_2$ and SiO$_2$ were measured to be 0.17, 0.022 and 0.058 \AA/s, respectively. The low deposition rates offer a high control in layer thickness and interfacial quality when depositing nanometer thick layers. The samples were designed to have a 6 nm top layer of Ta$_2$O$_5$ and 15 pairs of 1.6 nm SiO$_2$ or TiO$_2$ and 6 nm Ta$_2$O$_5$ (Figure 1). The stacks were deposited onto fused silica and silicon substrates by sequentially biasing the corresponding metal targets. For these designs, the cation concentration Si/(Si+Ta) and Ti/(Ti+Ta) is $\sim$20\%.  Reference mixture samples of the same cation concentration were also prepared. The mixture cation composition is controlled by varying the target bias time in one period.

\begin{figure}[ht]
\centering
  \includegraphics[height=8.5cm]{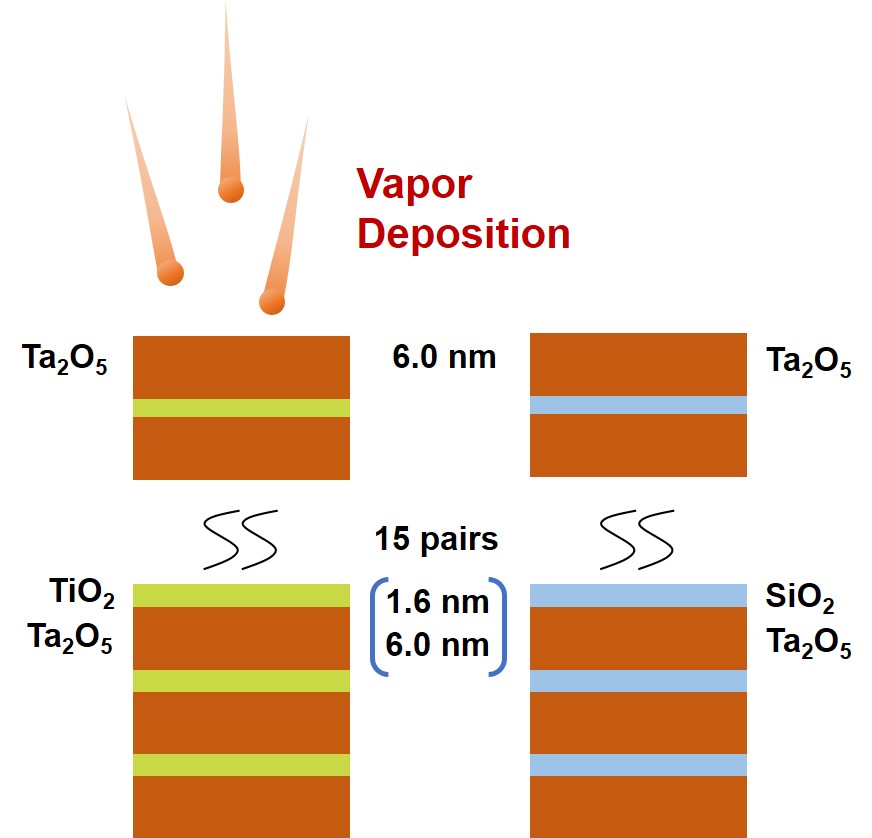}
  \caption{Schematic of the nanolaminate samples with a 6 nm top layer of Ta$_2$O$_5$ and 15 pairs of 1.6 nm SiO$_2$ or TiO$_2$ and 6 nm Ta$_2$O$_5$.}
   \label{fig1}
\end{figure}

The nanolaminate and mixture samples were annealed in air by ramping up the temperature at 1.6$^{\circ}$C/min and soaking at the set temperature for 10 hrs with a Fisher Scientific Isotemp programmable furnace. Soaking temperatures were varied in steps of 100$^{\circ}$C up to the highest temperature at which the first sign of crystallization appeared.

The structure evolution with annealing was characterized by grazing incidence x-ray diffraction (GIXRD) using a Bruker D8 thin film diffractometer operated at an incident angle of 0.5$^{\circ}$. X-ray photoelectron spectroscopy (XPS) measurements were carried out with a PE-5800 to examine the bonding environments of the elements. A take-off angle of 45$^\circ$ was used for all scans. The neutralizer operating at 10 $\mu$A was used to counteract charging effect of the sample. The C 1\textit{s} peak position was used to calibrate the binding energy scale of the spectra. The peak shapes were fitted with Gaussian functions. High resolution transmission electron microscopy images of as-deposited and annealed nanolaminates were obtained using a FEI Tecnai Osiris FEG/TEM operated at 200 kV by EAG Laboratories. 
 
The evaluation of the internal friction for each sample was performed with a coating ring-down system \cite{vajente2017high,cesarini2009gentle,granata2020progress} by Vajente et al at the LIGO laboratory, Caltech. A gentle nodal suspension was used to support the sample within a vacuum chamber with a pressure below 10$^{-6}$ Torr. After exciting the resonant mode, the decay of the oscillation amplitude was measured for each sample to obtain the internal friction.

\section{Results and discussion}
 
 Figure 2 shows the room temperature internal friction of the SiO$_2$/Ta$_2$O$_5$ and TiO$_2$/Ta$_2$O$_5$ nanolaminates for different annealing temperatures. For as-deposited nanolaminates, TiO$_2$/Ta$_2$O$_5$ has a higher Q$^{-1}$ $\approx$ (7.5 $\pm$ 0.2) $\times$ 10$^{-4}$ than SiO$_2$/Ta$_2$O$_5$ with Q$^{-1}$ $\approx$ (6.9 $\pm$ 0.5) $\times$ 10$^{-4}$. This is due to as-deposited SiO$_2$ single layer having low Q$^{-1}$ \cite{flaminio2010study}. Upon increasing the annealing temperature from 400$^\circ$C to 500$^\circ$C, the trend is reversed in that Q$^{-1}$ of the TiO$_2$/Ta$_2$O$_5$ is lower than that of the SiO$_2$/Ta$_2$O$_5$ nanolaminate. At 650$^\circ$C, the internal friction of the TiO$_2$/Ta$_2$O$_5$ nanolaminate reaches its lowest value of  Q$^{-1}$ $\approx$ (2.6 $\pm$ 0.2) $\times$ 10$^{-4}$. At this temperature,  Q$^{-1}$ is (3.1 $\pm$ 0.2) $\times$ 10$^{-4}$ for the SiO$_2$/Ta$_2$O$_5$ nanolaminate. In both cases, these values are lower than the best value for a single layer Ta$_2$O$_5$ annealed at 500$^\circ$C, Q$^{-1}$ $\approx$ 4.0 $\times$ 10$^{-4}$\cite{yang2019investigation}. For comparison, the inset of Figure 2 plots internal friction of the nanolaminates and mixtures of SiO$_2$/Ta$_2$O$_5$ and TiO$_2$/Ta$_2$O$_5$ at the highest annealing temperature before crystallization. For the SiO$_2$/Ta$_2$O$_5$ nanolaminate, Q$^{-1}$ is lower than that of the reference mixture of Q$^{-1}$ $\approx$ 4.1 $\times$ 10$^{-4}$ after 700$^\circ$C annealing. Phase separation after annealing in the SiO$_2$/Ta$_2$O$_5$ mixture results in a Q$^{-1}$ value similar to that of a single layer Ta$_2$O$_5$ (Figure S1). Instead, Q$^{-1}$ is the same within experimental errors for the TiO$_2$/Ta$_2$O$_5$ nanolaminate and mixture after annealing, indicating a strong similarity. In both the mixture and nanolaminate form, the TiO$_2$/Ta$_2$O$_5$ system exhibits lower internal friction than that of the SiO$_2$/Ta$_2$O$_5$ one.
 
 \begin{figure}[ht]
     \centering
     \includegraphics[height=5.8cm]{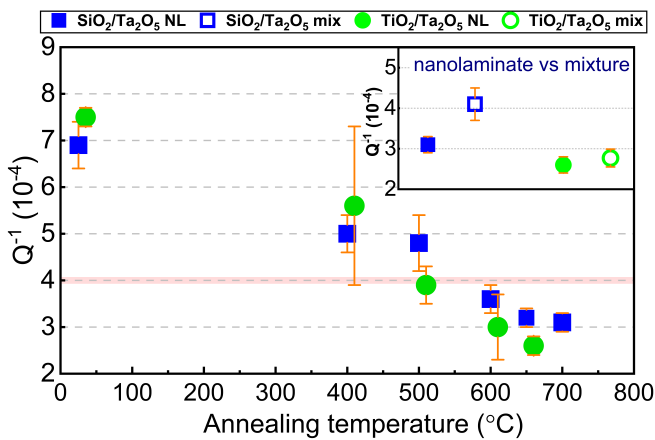}
     \caption{Room temperature internal friction of nanolaminates as-deposited and annealed at different temperatures. The SiO$_2$/Ta$_2$O$_5$ nanolaminate (NL) is represented by blue filled squares and the TiO$_2$/Ta$_2$O$_5$ nanolaminate (NL) is represented by green filled circles. After annealing at 500$^\circ$C, the TiO$_2$/Ta$_2$O$_5$ nanolaminate has a lower internal friction than the SiO$_2$/Ta$_2$O$_5$ nanolaminate. Inset, internal friction of the SiO$_2$/Ta$_2$O$_5$ nanolaminate annealed at 700$^\circ$C, the SiO$_2$/Ta$_2$O$_5$ mixture annealed at 700$^\circ$C (blue open square), the TiO$_2$/Ta$_2$O$_5$ nanolaminate annealed at 650$^\circ$C, and the TiO$_2$/Ta$_2$O$_5$ mixture annealed at 600$^\circ$C (green open circle). The red shading indicates the internal friction level of a single layer Ta$_2$O$_5$ annealed at 500$^\circ$C.}
     \label{fig2}
 \end{figure}

 There are significant differences in the structural evolution upon annealing that affect the internal friction behavior of the nanolaminates. The high resolution TEM images of the as-deposited TiO$_2$/Ta$_2$O$_5$ and SiO$_2$/Ta$_2$O$_5$ nanolaminates show well defined layers with sharp interfaces (Figure 3). Upon annealing at 650$^\circ$C, the SiO$_2$/Ta$_2$O$_5$ nanolaminate remains unchanged, showing intact interfaces between strongly contrasted SiO$_2$ and Ta$_2$O$_5$ layers. In this case, the Q$^{-1}$ can be approximated as the weighted average of the internal friction of the composing oxide layers (disregarding any interfacial effects)\cite{kuo2019low, granata2018correlated}. This calculation estimates Q$^{-1}$ $\approx$ 3.4 $\times$ 10$^{-4}$, which is in good agreement with the measured value. On the contrary, a noticeable discontinuity along the interfaces and a homogenization of the two oxide materials are found in the annealed TiO$_2$/Ta$_2$O$_5$ nanolaminate.

 \begin{figure}[ht]
     \centering
     \includegraphics[height=8cm]{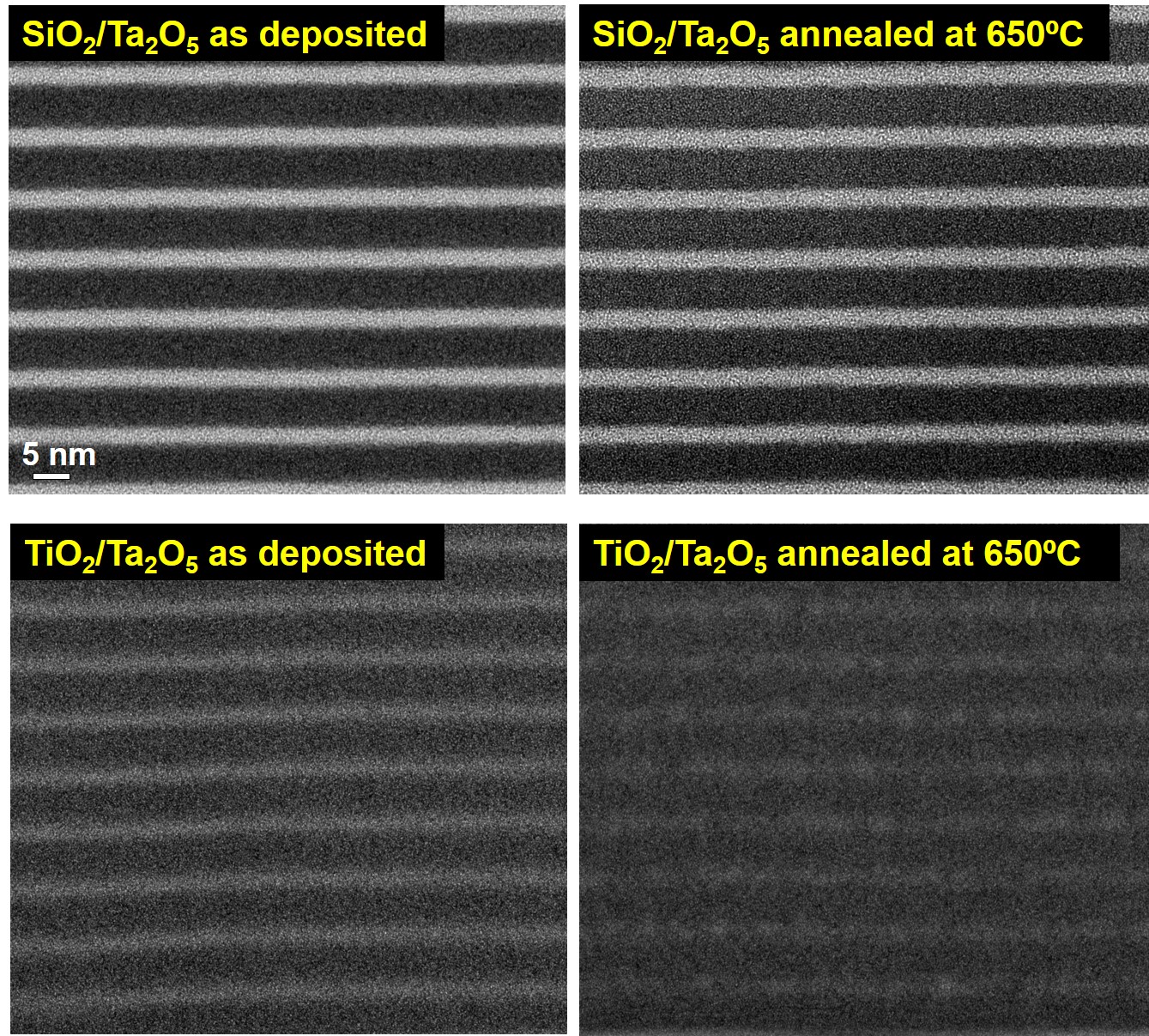}
     \caption{High resolution cross-sectional TEM images of nanolaminates before and after annealing at 650$^\circ$C. Top,  SiO$_2$/Ta$_2$O$_5$ nanolaminate showing robust layer structure before and after annealing. Bottom,TiO$_2$/Ta$_2$O$_5$ nanolaminate showing interface blurring after annealing.}
     \label{fig3}
 \end{figure}

Full crystallization of the TiO$_2$/Ta$_2$O$_5$ and SiO$_2$/Ta$_2$O$_5$ nanolaminates is observed after annealing at 750$^\circ$C and 800$^\circ$C, respectively. The crystallization process is delayed to a higher annealing temperature in both nanolaminates compared to 675$^\circ$C, the crystallization temperature of a single layer Ta$_2$O$_5$ \cite{yang2019investigation}, due to a greater contribution from the surface enthalpy to the total energy \cite{pan2014thickness,zhang2017thickness,wang2009atomic}. Diffraction patterns of the crystallized nanolaminates and the crystallized Ta$_2$O$_5$ single layer are shown in Figure 4. The spectrum of the crystallized SiO$_2$/Ta$_2$O$_5$ nanolaminate only exhibits diffraction peaks from the orthorhombic Ta$_2$O$_5$ phase \cite{chaneliere1998properties} while no SiO$_2$ peak is found. The peaks feature a much broader profile than that of a single layer Ta$_2$O$_5$. Applying the Scherrer equation \cite{langford1978scherrer} to the peak at 2theta = 28.7 $^\circ$, the crystallite size in the SiO$_2$/Ta$_2$O$_5$ nanolaminate is calculated to be $\sim$ 5 nm, while a size of 35 nm is calculated for the single layer Ta$_2$O$_5$. The crystallite size of 5 nm is fully consistent with the Ta$_2$O$_5$ layer thickness $\sim$ 6 nm, indicating the physical constraint to crystallite growth imposed by the interfaces. The same analysis of the TiO$_2$/Ta$_2$O$_5$ nanolaminate yields a crystallite size of 11 nm that is larger than the layer thickness. The extended growth is only possible when bi-directional diffusion across the interface bridges the separated Ta$_2$O$_5$ layers. 

\begin{figure}[ht]
     \centering
     \includegraphics[height=5cm]{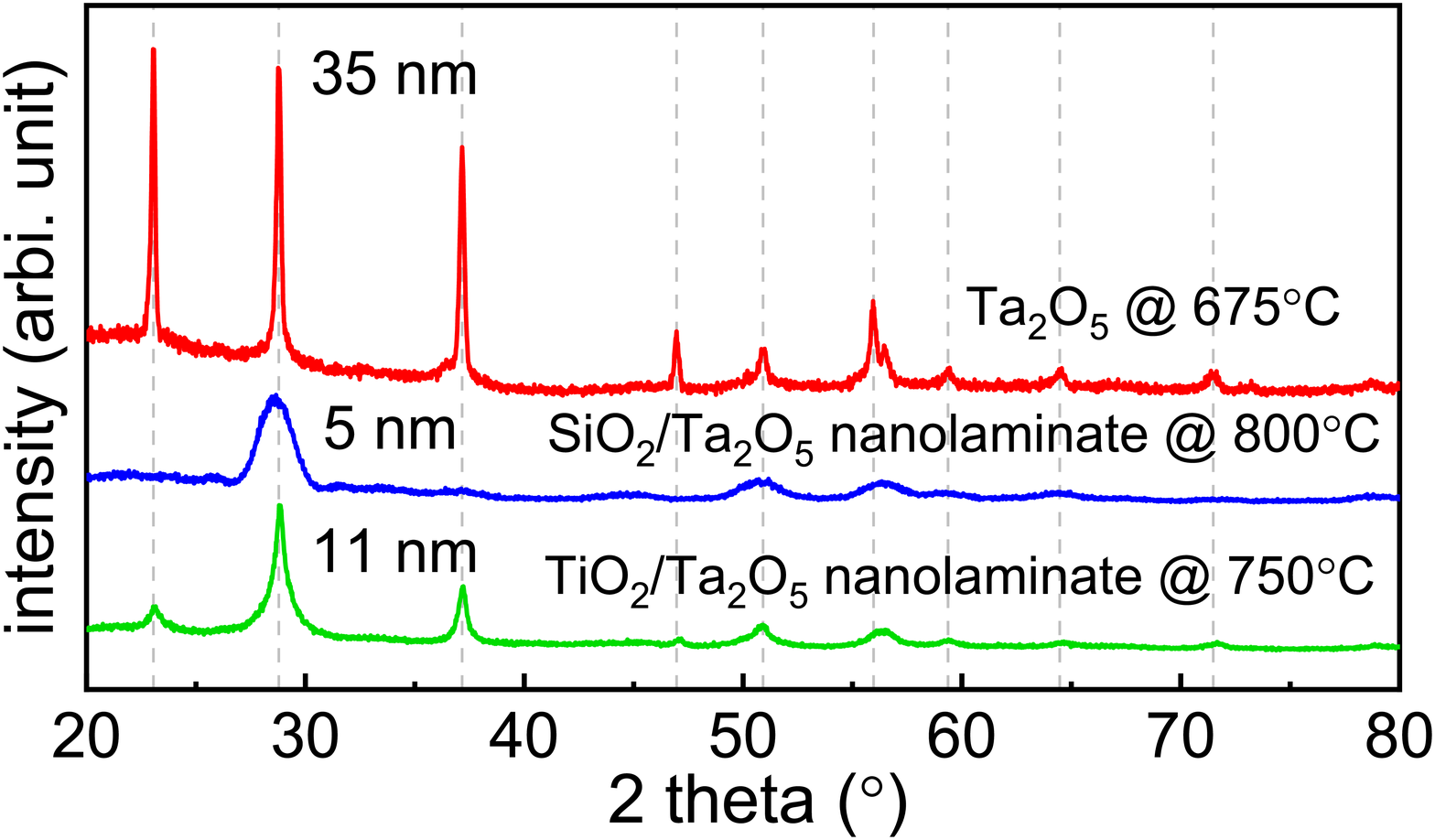}
     \caption{Diffraction patterns of the SiO$_2$/Ta$_2$O$_5$ nanolaminate annealed at 800$^\circ$C (blue line) and the TiO$_2$/Ta$_2$O$_5$ nanolaminate annealed at 750$^\circ$C (green line) after crystallization. For comparison, the diffraction spectrum of a crystallized Ta$_2$O$_5$ single layer annealed at 675$^\circ$C is shown on top (red line).}
     \label{fig4}
 \end{figure}

The emergence of Ti 2p peaks in the XPS spectrum confirms a strong Ti cation diffusion into the top Ta$_2$O$_5$ layer in the TiO$_2$/Ta$_2$O$_5$ nanolaminate upon annealing (Figure S2). Instead, a Si 2p peak is almost absent in the XPS spectrum of both as-deposited and annealed SiO$_2$/Ta$_2$O$_5$ nanolaminates. Figure 5 shows the Ti 2p core level in the TiO$_2$/Ta$_2$O$_5$ nanolaminate after annealing at 650$^\circ$C. The peak separation is determined to be $\sim$ 5.85 $\pm$ 0.02 eV, which is significantly different from 5.7 eV for Ti in a TiO$_2$(IV) environment \cite{barreca2007tio2}. The change in peak separation is ascribed to the Ta-O-Ti bonding process that accompanies the mixing of TiO$_2$ and Ta$_2$O$_5$. The mixing in this sample is the same as observed in a TiO$_2$/Ta$_2$O$_5$ mixture upon annealing, which leads to the formation of a ternary compound identified as  TiTa$_{18}$O$_{47}$ that is thermodynamically more stable than separate phases of TiO$_2$ and Ta$_2$O$_5$ \cite{fazio2020structure}. The existence of a stable ternary phase in nanolaminates has been well demonstrated in systems such as Al$_2$O$_3$/TiO$_2$ \cite{kim2014al2o3,kim2016optimization} and Al$_2$O$_3$/ZrO$_2$ \cite{meyer2009al2o3,meyer2010origin}, where the presence of the mixed phase gives superior physical and/or chemical properties that can be used in various applications. In the absence of a stable ternary phase, a cross-interface diffusion between layers should not occur as it will not lead to a reduction in Gibbs free energy for the whole system. Such picture is supported by the observation of well spaced layers in the SiO$_2$/Ta$_2$O$_5$ (Figure 3) and SiO$_2$/TiO$_2$ \cite{kuo2019low} nanolaminates after annealing .

\begin{figure}[ht]
     \centering
     \includegraphics[height=5cm]{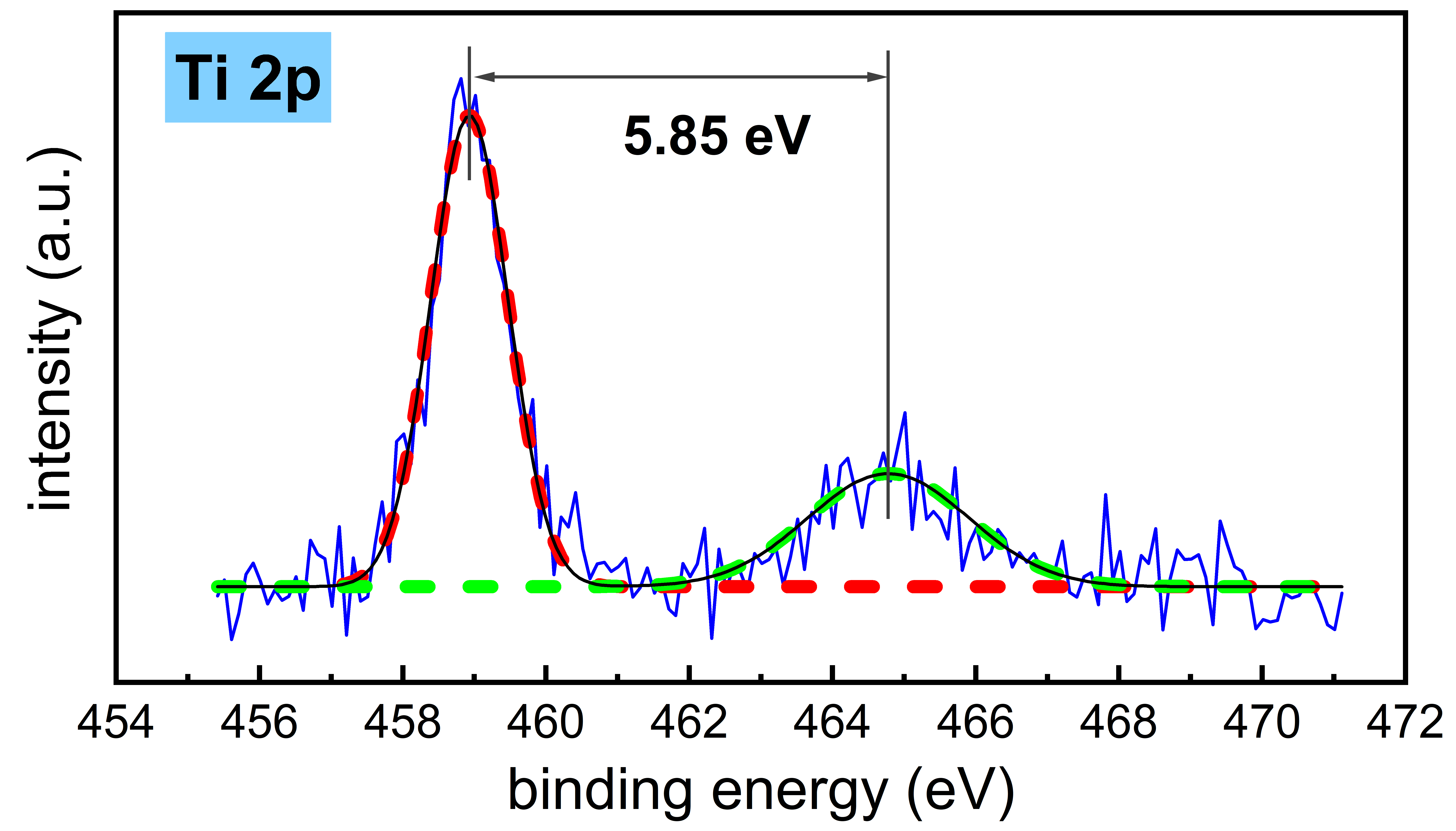}
     \caption{X-ray photoelectron spectrum of Ti 2p in the TiO$_2$/Ta$_2$O$_5$ nanolaminate after annealing at 650$^\circ$C. The peak separation is determined to be 5.85 eV. Collected spectrum after background subtraction is shown in blue line, the composite spectrum is shown in black line, the Ti 2p 3/2 peak is shown in red dashed line, and  the Ti 2p 1/2 peak is shown in green dashed line.}
     \label{fig5}
 \end{figure}

These results show that in these binary oxide nanolaminates, the more significant reduction in room temperature internal friction is dominated by interdiffusion and the formation of a stable mixed phase rather than the effect of nanolayering of the two oxide materials. For the TiO$_2$/Ta$_2$O$_5$ nanolaminate, it is the reorganization that occurs during the mixing that has a major influence on reducing the room temperature internal friction. Elevated annealing temperature in the SiO$_2$/Ta$_2$O$_5$ nanolaminate improves the internal friction with respect to a single layer Ta$_2$O$_5$ due to a reduction of internal friction in both the SiO$_2$ and Ta$_2$O$_5$ nanolayers. The internal friction of an annealed SiO$_2$ layer reaches Q$^{-1}$ $\approx$ 0.5 $\times$ 10$^{-4}$ that is the lowest value among all binary oxide films \cite{flaminio2010study}. Yet the reduction is not as profound as from mixing in the TiO$_2$/Ta$_2$O$_5$ nanolaminate. This behavior is different from what has been observed for the cryogenic internal friction of SiO$_2$/TiO$_2$ nanolaminates, which manifests itself by a decrease in internal friction with a decrease in layer thickness \cite{kuo2019low}. Such effect is ascribed to the elimination of two-level system transitions with characteristic dimensions that exceed the layer thickness. At room temperature, however, contribution from those cryogenic two-level systems to the internal friction is negligibly small \cite{jiang2020atomic} so that the effect from nanolayering does not lead to lower internal friction. On the contrary, observation of the structural evolution in the TiO$_2$/Ta$_2$O$_5$ nanolaminate with annealing suggests that the formation of a more stable phase might result in the modified distribution of two-level systems that is responsible for lowering the room temperature internal friction. 

\section{Conclusions}

An unambiguous identification of the morphological changes that involve layer breakup and mixture formation in the TiO$_2$/Ta$_2$O$_5$ nanolaminate is demonstrated. The mixing of TiO$_2$ and Ta$_2$O$_5$, which is thermodynamically favored upon annealing, results in the major improvements in the room temperature internal friction of the TiO$_2$/Ta$_2$O$_5$ nanolaminate. In the absence of mixing, the reduction in room temperature internal friction of a nanolaminate with annealing can be predicted from the changes in the composing layers, as is the case of the SiO$_2$/Ta$_2$O$_5$ nanolaminate. A modified distribution of two-level systems in the mixed phase rather than size effects is argued to be responsible for the more profoundly reduced room temperature internal friction in the TiO$_2$/Ta$_2$O$_5$ nanolaminate. These results are significant in that they provide new insight into the suppression of two-level systems and pave the way towards a more delineated search of new coating materials with low internal friction that potentially will increase the sensitivity of the next generation gravitational wave detectors.

\section*{Conflicts of interest}
There are no conflicts to declare.

\section*{Acknowledgements}

This work is supported by the National Science Foundation LIGO program through grants No. 1710957 and 1708010. We also acknowledge the support of the LSC Center for Coatings Research, jointly funded by the National Science Foundation (NSF) and the Gordon and Betty Moore Foundation. A. M., R. B. and M. M. F. are grateful for support through NSF awards PHY-1707866, PHY- 1708175 and GBMF Grant No. 6793. Partial support from grant ONR No. N00014-17-1-2536 is acknowledged. We thank Riccardo DeSalvo, Innocenzo Pinto, and Manel Molina Ruiz for useful discussions.

%%%END OF MAIN TEXT%%%


\balance


%%%REFERENCES%%%
\bibliography{rsc} %You need to replace "rsc" on this line with the name of your .bib file

\providecommand*{\mcitethebibliography}{\thebibliography}
\csname @ifundefined\endcsname{endmcitethebibliography}
{\let\endmcitethebibliography\endthebibliography}{}
\begin{mcitethebibliography}{35}
\providecommand*{\natexlab}[1]{#1}
\providecommand*{\mciteSetBstSublistMode}[1]{}
\providecommand*{\mciteSetBstMaxWidthForm}[2]{}
\providecommand*{\mciteBstWouldAddEndPuncttrue}
  {\def\EndOfBibitem{\unskip.}}
\providecommand*{\mciteBstWouldAddEndPunctfalse}
  {\let\EndOfBibitem\relax}
\providecommand*{\mciteSetBstMidEndSepPunct}[3]{}
\providecommand*{\mciteSetBstSublistLabelBeginEnd}[3]{}
\providecommand*{\EndOfBibitem}{}
\mciteSetBstSublistMode{f}
\mciteSetBstMaxWidthForm{subitem}
{(\emph{\alph{mcitesubitemcount}})}
\mciteSetBstSublistLabelBeginEnd{\mcitemaxwidthsubitemform\space}
{\relax}{\relax}

\bibitem[Tsipas \emph{et~al.}(2008)Tsipas, Volkos, Sotiropoulos, Galata,
  Mavrou, Tsoutsou, Panayiotatos, Dimoulas, Marchiori, and
  Fompeyrine]{tsipas2008germanium}
P.~Tsipas, S.~Volkos, A.~Sotiropoulos, S.~Galata, G.~Mavrou, D.~Tsoutsou,
  Y.~Panayiotatos, A.~Dimoulas, C.~Marchiori and J.~Fompeyrine, \emph{Applied
  physics letters}, 2008, \textbf{93}, 082904\relax
\mciteBstWouldAddEndPuncttrue
\mciteSetBstMidEndSepPunct{\mcitedefaultmidpunct}
{\mcitedefaultendpunct}{\mcitedefaultseppunct}\relax
\EndOfBibitem
\bibitem[Wilk \emph{et~al.}(2000)Wilk, Wallace, and Anthony]{wilk2000hafnium}
G.~Wilk, R.~Wallace and J.~Anthony, \emph{Journal of Applied Physics}, 2000,
  \textbf{87}, 484--492\relax
\mciteBstWouldAddEndPuncttrue
\mciteSetBstMidEndSepPunct{\mcitedefaultmidpunct}
{\mcitedefaultendpunct}{\mcitedefaultseppunct}\relax
\EndOfBibitem
\bibitem[Boscherini \emph{et~al.}(2011)Boscherini, D’Acapito, Galata,
  Tsoutsou, and Dimoulas]{boscherini2011atomic}
F.~Boscherini, F.~D’Acapito, S.~Galata, D.~Tsoutsou and A.~Dimoulas,
  \emph{Applied Physics Letters}, 2011, \textbf{99}, 121909\relax
\mciteBstWouldAddEndPuncttrue
\mciteSetBstMidEndSepPunct{\mcitedefaultmidpunct}
{\mcitedefaultendpunct}{\mcitedefaultseppunct}\relax
\EndOfBibitem
\bibitem[Huo \emph{et~al.}(2020)Huo, Wang, Li, Liu, and Li]{huo2020two}
X.~Huo, X.~Wang, Z.~Li, J.~Liu and J.~Li, \emph{Nanoscale}, 2020, \textbf{12},
  3387--3399\relax
\mciteBstWouldAddEndPuncttrue
\mciteSetBstMidEndSepPunct{\mcitedefaultmidpunct}
{\mcitedefaultendpunct}{\mcitedefaultseppunct}\relax
\EndOfBibitem
\bibitem[Sahoo and Shapiro(1998)]{sahoo1998mgo}
N.~K. Sahoo and A.~P. Shapiro, \emph{Applied optics}, 1998, \textbf{37},
  8043--8056\relax
\mciteBstWouldAddEndPuncttrue
\mciteSetBstMidEndSepPunct{\mcitedefaultmidpunct}
{\mcitedefaultendpunct}{\mcitedefaultseppunct}\relax
\EndOfBibitem
\bibitem[Karvonen \emph{et~al.}(2013)Karvonen, S{\"a}yn{\"a}tjoki, Chen,
  Jussila, R{\"o}nn, Ruoho, Alasaarela, Kujala, Norwood,
  Peyghambarian,\emph{et~al.}]{karvonen2013enhancement}
L.~Karvonen, A.~S{\"a}yn{\"a}tjoki, Y.~Chen, H.~Jussila, J.~R{\"o}nn, M.~Ruoho,
  T.~Alasaarela, S.~Kujala, R.~A. Norwood, N.~Peyghambarian \emph{et~al.},
  \emph{Applied Physics Letters}, 2013, \textbf{103}, 031903\relax
\mciteBstWouldAddEndPuncttrue
\mciteSetBstMidEndSepPunct{\mcitedefaultmidpunct}
{\mcitedefaultendpunct}{\mcitedefaultseppunct}\relax
\EndOfBibitem
\bibitem[Li and Bi(2017)]{li2017quantum}
J.~Li and X.~Bi, \emph{Nanoscale}, 2017, \textbf{9}, 16420--16428\relax
\mciteBstWouldAddEndPuncttrue
\mciteSetBstMidEndSepPunct{\mcitedefaultmidpunct}
{\mcitedefaultendpunct}{\mcitedefaultseppunct}\relax
\EndOfBibitem
\bibitem[Wang and Ekerdt(2011)]{wang2011structure}
T.~Wang and J.~G. Ekerdt, \emph{Chemistry of Materials}, 2011, \textbf{23},
  1679--1685\relax
\mciteBstWouldAddEndPuncttrue
\mciteSetBstMidEndSepPunct{\mcitedefaultmidpunct}
{\mcitedefaultendpunct}{\mcitedefaultseppunct}\relax
\EndOfBibitem
\bibitem[Wang and Ekerdt(2009)]{wang2009atomic}
T.~Wang and J.~G. Ekerdt, \emph{Chemistry of Materials}, 2009, \textbf{21},
  3096--3101\relax
\mciteBstWouldAddEndPuncttrue
\mciteSetBstMidEndSepPunct{\mcitedefaultmidpunct}
{\mcitedefaultendpunct}{\mcitedefaultseppunct}\relax
\EndOfBibitem
\bibitem[Kim \emph{et~al.}(2014)Kim, Kim, Park, Jeong, Kim, Chung, Kim, and
  Park]{kim2014al2o3}
L.~H. Kim, K.~Kim, S.~Park, Y.~J. Jeong, H.~Kim, D.~S. Chung, S.~H. Kim and
  C.~E. Park, \emph{ACS Applied Materials \& Interfaces}, 2014, \textbf{6},
  6731--6738\relax
\mciteBstWouldAddEndPuncttrue
\mciteSetBstMidEndSepPunct{\mcitedefaultmidpunct}
{\mcitedefaultendpunct}{\mcitedefaultseppunct}\relax
\EndOfBibitem
\bibitem[Kim \emph{et~al.}(2016)Kim, Jeong, An, Park, Jang, Nam, Jang, Kim, and
  Park]{kim2016optimization}
L.~H. Kim, Y.~J. Jeong, T.~K. An, S.~Park, J.~H. Jang, S.~Nam, J.~Jang, S.~H.
  Kim and C.~E. Park, \emph{Physical Chemistry Chemical Physics}, 2016,
  \textbf{18}, 1042--1049\relax
\mciteBstWouldAddEndPuncttrue
\mciteSetBstMidEndSepPunct{\mcitedefaultmidpunct}
{\mcitedefaultendpunct}{\mcitedefaultseppunct}\relax
\EndOfBibitem
\bibitem[Meyer \emph{et~al.}(2009)Meyer, G{\"o}rrn, Bertram, Hamwi, Winkler,
  Johannes, Weimann, Hinze, Riedl, and Kowalsky]{meyer2009al2o3}
J.~Meyer, P.~G{\"o}rrn, F.~Bertram, S.~Hamwi, T.~Winkler, H.-H. Johannes,
  T.~Weimann, P.~Hinze, T.~Riedl and W.~Kowalsky, \emph{Advanced materials},
  2009, \textbf{21}, 1845--1849\relax
\mciteBstWouldAddEndPuncttrue
\mciteSetBstMidEndSepPunct{\mcitedefaultmidpunct}
{\mcitedefaultendpunct}{\mcitedefaultseppunct}\relax
\EndOfBibitem
\bibitem[Zhao \emph{et~al.}(2002)Zhao, Richard, Bender, Caymax, De~Gendt,
  Heyns, Young, Roebben, Van~der Biest, and Haukka]{zhao2002miscibility}
C.~Zhao, O.~Richard, H.~Bender, M.~Caymax, S.~De~Gendt, M.~Heyns, E.~Young,
  G.~Roebben, O.~Van~der Biest and S.~Haukka, \emph{Applied physics letters},
  2002, \textbf{80}, 2374--2376\relax
\mciteBstWouldAddEndPuncttrue
\mciteSetBstMidEndSepPunct{\mcitedefaultmidpunct}
{\mcitedefaultendpunct}{\mcitedefaultseppunct}\relax
\EndOfBibitem
\bibitem[Meyer \emph{et~al.}(2010)Meyer, Schmidt, Kowalsky, Riedl, and
  Kahn]{meyer2010origin}
J.~Meyer, H.~Schmidt, W.~Kowalsky, T.~Riedl and A.~Kahn, \emph{Applied Physics
  Letters}, 2010, \textbf{96}, 117\relax
\mciteBstWouldAddEndPuncttrue
\mciteSetBstMidEndSepPunct{\mcitedefaultmidpunct}
{\mcitedefaultendpunct}{\mcitedefaultseppunct}\relax
\EndOfBibitem
\bibitem[Cole \emph{et~al.}(2013)Cole, Zhang, Martin, Ye, and
  Aspelmeyer]{cole2013tenfold}
G.~D. Cole, W.~Zhang, M.~J. Martin, J.~Ye and M.~Aspelmeyer, \emph{Nature
  Photonics}, 2013, \textbf{7}, 644--650\relax
\mciteBstWouldAddEndPuncttrue
\mciteSetBstMidEndSepPunct{\mcitedefaultmidpunct}
{\mcitedefaultendpunct}{\mcitedefaultseppunct}\relax
\EndOfBibitem
\bibitem[Morell \emph{et~al.}(2016)Morell, Reserbat-Plantey, Tsioutsios,
  Sch{\"a}dler, Dubin, Koppens, and Bachtold]{morell2016high}
N.~Morell, A.~Reserbat-Plantey, I.~Tsioutsios, K.~G. Sch{\"a}dler, F.~Dubin,
  F.~H. Koppens and A.~Bachtold, \emph{Nano letters}, 2016, \textbf{16},
  5102--5108\relax
\mciteBstWouldAddEndPuncttrue
\mciteSetBstMidEndSepPunct{\mcitedefaultmidpunct}
{\mcitedefaultendpunct}{\mcitedefaultseppunct}\relax
\EndOfBibitem
\bibitem[Miller \emph{et~al.}(2015)Miller, Barsotti, Vitale, Fritschel, Evans,
  and Sigg]{miller2015prospects}
J.~Miller, L.~Barsotti, S.~Vitale, P.~Fritschel, M.~Evans and D.~Sigg,
  \emph{Physical Review D}, 2015, \textbf{91}, 062005\relax
\mciteBstWouldAddEndPuncttrue
\mciteSetBstMidEndSepPunct{\mcitedefaultmidpunct}
{\mcitedefaultendpunct}{\mcitedefaultseppunct}\relax
\EndOfBibitem
\bibitem[Vajente \emph{et~al.}(2019)Vajente, Gustafson, and
  Reitze]{vajente2019precision}
G.~Vajente, E.~K. Gustafson and D.~H. Reitze, \emph{Advances In Atomic,
  Molecular, and Optical Physics}, Elsevier, 2019, vol.~68, pp. 75--148\relax
\mciteBstWouldAddEndPuncttrue
\mciteSetBstMidEndSepPunct{\mcitedefaultmidpunct}
{\mcitedefaultendpunct}{\mcitedefaultseppunct}\relax
\EndOfBibitem
\bibitem[Abbott \emph{et~al.}(2018)Abbott, Abbott, Abbott, Abernathy, Acernese,
  Ackley, Adams, Adams, Addesso, Adhikari,\emph{et~al.}]{abbott2018prospects}
B.~P. Abbott, R.~Abbott, T.~Abbott, M.~Abernathy, F.~Acernese, K.~Ackley,
  C.~Adams, T.~Adams, P.~Addesso, R.~Adhikari \emph{et~al.}, \emph{Living
  Reviews in Relativity}, 2018, \textbf{21}, 3\relax
\mciteBstWouldAddEndPuncttrue
\mciteSetBstMidEndSepPunct{\mcitedefaultmidpunct}
{\mcitedefaultendpunct}{\mcitedefaultseppunct}\relax
\EndOfBibitem
\bibitem[Billman \emph{et~al.}(2017)Billman, Trinastic, Davis, Hamdan, and
  Cheng]{billman2017origin}
C.~R. Billman, J.~P. Trinastic, D.~J. Davis, R.~Hamdan and H.-P. Cheng,
  \emph{Physical Review B}, 2017, \textbf{95}, 014109\relax
\mciteBstWouldAddEndPuncttrue
\mciteSetBstMidEndSepPunct{\mcitedefaultmidpunct}
{\mcitedefaultendpunct}{\mcitedefaultseppunct}\relax
\EndOfBibitem
\bibitem[Kuo \emph{et~al.}(2019)Kuo, Pan, Chang, and Chao]{kuo2019low}
L.-C. Kuo, H.-W. Pan, C.-L. Chang and S.~Chao, \emph{Optics letters}, 2019,
  \textbf{44}, 247--250\relax
\mciteBstWouldAddEndPuncttrue
\mciteSetBstMidEndSepPunct{\mcitedefaultmidpunct}
{\mcitedefaultendpunct}{\mcitedefaultseppunct}\relax
\EndOfBibitem
\bibitem[West \emph{et~al.}(2008)West, Lu, Yu, Kirkwood, Chen, Pei, Claassen,
  and Wolf]{west2008growth}
K.~G. West, J.~Lu, J.~Yu, D.~Kirkwood, W.~Chen, Y.~Pei, J.~Claassen and S.~A.
  Wolf, \emph{Journal of Vacuum Science \& Technology A: Vacuum, Surfaces, and
  Films}, 2008, \textbf{26}, 133--139\relax
\mciteBstWouldAddEndPuncttrue
\mciteSetBstMidEndSepPunct{\mcitedefaultmidpunct}
{\mcitedefaultendpunct}{\mcitedefaultseppunct}\relax
\EndOfBibitem
\bibitem[Vajente \emph{et~al.}(2017)Vajente, Ananyeva, Billingsley, Gustafson,
  Heptonstall, Sanchez, and Torrie]{vajente2017high}
G.~Vajente, A.~Ananyeva, G.~Billingsley, E.~Gustafson, A.~Heptonstall,
  E.~Sanchez and C.~Torrie, \emph{Review of Scientific Instruments}, 2017,
  \textbf{88}, 073901\relax
\mciteBstWouldAddEndPuncttrue
\mciteSetBstMidEndSepPunct{\mcitedefaultmidpunct}
{\mcitedefaultendpunct}{\mcitedefaultseppunct}\relax
\EndOfBibitem
\bibitem[Cesarini \emph{et~al.}(2009)Cesarini, Lorenzini, Campagna, Martelli,
  Piergiovanni, Vetrano, Losurdo, and Cagnoli]{cesarini2009gentle}
E.~Cesarini, M.~Lorenzini, E.~Campagna, F.~Martelli, F.~Piergiovanni,
  F.~Vetrano, G.~Losurdo and G.~Cagnoli, \emph{Review of Scientific
  Instruments}, 2009, \textbf{80}, 053904\relax
\mciteBstWouldAddEndPuncttrue
\mciteSetBstMidEndSepPunct{\mcitedefaultmidpunct}
{\mcitedefaultendpunct}{\mcitedefaultseppunct}\relax
\EndOfBibitem
\bibitem[Granata \emph{et~al.}(2020)Granata, Amato, Cagnoli, Coulon, Degallaix,
  Forest, Mereni, Michel, Pinard, Sassolas,\emph{et~al.}]{granata2020progress}
M.~Granata, A.~Amato, G.~Cagnoli, M.~Coulon, J.~Degallaix, D.~Forest,
  L.~Mereni, C.~Michel, L.~Pinard, B.~Sassolas \emph{et~al.}, \emph{Applied
  Optics}, 2020, \textbf{59}, A229--A235\relax
\mciteBstWouldAddEndPuncttrue
\mciteSetBstMidEndSepPunct{\mcitedefaultmidpunct}
{\mcitedefaultendpunct}{\mcitedefaultseppunct}\relax
\EndOfBibitem
\bibitem[Flaminio \emph{et~al.}(2010)Flaminio, Franc, Michel, Morgado, Pinard,
  and Sassolas]{flaminio2010study}
R.~Flaminio, J.~Franc, C.~Michel, N.~Morgado, L.~Pinard and B.~Sassolas,
  \emph{Classical and Quantum Gravity}, 2010, \textbf{27}, 084030\relax
\mciteBstWouldAddEndPuncttrue
\mciteSetBstMidEndSepPunct{\mcitedefaultmidpunct}
{\mcitedefaultendpunct}{\mcitedefaultseppunct}\relax
\EndOfBibitem
\bibitem[Yang \emph{et~al.}(2019)Yang, Randel, Vajente, Ananyeva, Gustafson,
  Markosyan, Bassiri, Fejer, and Menoni]{yang2019investigation}
L.~Yang, E.~Randel, G.~Vajente, A.~Ananyeva, E.~Gustafson, A.~Markosyan,
  R.~Bassiri, M.~M. Fejer and C.~S. Menoni, \emph{Physical Review D}, 2019,
  \textbf{100}, 122004\relax
\mciteBstWouldAddEndPuncttrue
\mciteSetBstMidEndSepPunct{\mcitedefaultmidpunct}
{\mcitedefaultendpunct}{\mcitedefaultseppunct}\relax
\EndOfBibitem
\bibitem[Granata \emph{et~al.}(2018)Granata, Coillet, Martinez, Dolique, Amato,
  Canepa, Margueritat, Martinet, Mermet,
  Michel,\emph{et~al.}]{granata2018correlated}
M.~Granata, E.~Coillet, V.~Martinez, V.~Dolique, A.~Amato, M.~Canepa,
  J.~Margueritat, C.~Martinet, A.~Mermet, C.~Michel \emph{et~al.},
  \emph{Physical Review Materials}, 2018, \textbf{2}, 053607\relax
\mciteBstWouldAddEndPuncttrue
\mciteSetBstMidEndSepPunct{\mcitedefaultmidpunct}
{\mcitedefaultendpunct}{\mcitedefaultseppunct}\relax
\EndOfBibitem
\bibitem[Pan \emph{et~al.}(2014)Pan, Wang, Kuo, Chao, Principe, Pinto, and
  DeSalvo]{pan2014thickness}
H.-W. Pan, S.-J. Wang, L.-C. Kuo, S.~Chao, M.~Principe, I.~M. Pinto and
  R.~DeSalvo, \emph{Optics express}, 2014, \textbf{22}, 29847--29854\relax
\mciteBstWouldAddEndPuncttrue
\mciteSetBstMidEndSepPunct{\mcitedefaultmidpunct}
{\mcitedefaultendpunct}{\mcitedefaultseppunct}\relax
\EndOfBibitem
\bibitem[Zhang \emph{et~al.}(2017)Zhang, Zhang, Jiao, Bao, Wang, and
  Cheng]{zhang2017thickness}
L.~Zhang, J.~Zhang, H.~Jiao, G.~Bao, Z.~Wang and X.~Cheng, \emph{Thin Solid
  Films}, 2017, \textbf{642}, 359--363\relax
\mciteBstWouldAddEndPuncttrue
\mciteSetBstMidEndSepPunct{\mcitedefaultmidpunct}
{\mcitedefaultendpunct}{\mcitedefaultseppunct}\relax
\EndOfBibitem
\bibitem[Chaneliere \emph{et~al.}(1998)Chaneliere, Four, Autran, Devine, and
  Sandler]{chaneliere1998properties}
C.~Chaneliere, S.~Four, J.~Autran, R.~Devine and N.~Sandler, \emph{Journal of
  applied physics}, 1998, \textbf{83}, 4823--4829\relax
\mciteBstWouldAddEndPuncttrue
\mciteSetBstMidEndSepPunct{\mcitedefaultmidpunct}
{\mcitedefaultendpunct}{\mcitedefaultseppunct}\relax
\EndOfBibitem
\bibitem[Langford and Wilson(1978)]{langford1978scherrer}
J.~I. Langford and A.~Wilson, \emph{Journal of applied crystallography}, 1978,
  \textbf{11}, 102--113\relax
\mciteBstWouldAddEndPuncttrue
\mciteSetBstMidEndSepPunct{\mcitedefaultmidpunct}
{\mcitedefaultendpunct}{\mcitedefaultseppunct}\relax
\EndOfBibitem
\bibitem[Barreca \emph{et~al.}(2007)Barreca, Gasparotto, Maccato, Maragno, and
  Tondello]{barreca2007tio2}
D.~Barreca, A.~Gasparotto, C.~Maccato, C.~Maragno and E.~Tondello,
  \emph{Surface Science Spectra}, 2007, \textbf{14}, 27--33\relax
\mciteBstWouldAddEndPuncttrue
\mciteSetBstMidEndSepPunct{\mcitedefaultmidpunct}
{\mcitedefaultendpunct}{\mcitedefaultseppunct}\relax
\EndOfBibitem
\bibitem[Fazio \emph{et~al.}(2020)Fazio, Vajente, Ananyeva, Markosyan, Bassiri,
  Fejer, and Menoni]{fazio2020structure}
M.~Fazio, G.~Vajente, A.~Ananyeva, A.~Markosyan, R.~Bassiri, M.~Fejer and C.~S.
  Menoni, \emph{Opt. Mater. Express}, 2020, \textbf{10}, 1687\relax
\mciteBstWouldAddEndPuncttrue
\mciteSetBstMidEndSepPunct{\mcitedefaultmidpunct}
{\mcitedefaultendpunct}{\mcitedefaultseppunct}\relax
\EndOfBibitem
\bibitem[Jiang \emph{et~al.}()Jiang, Mishkin, Prasai, Yazback, Zhang, Bassiri,
  Fejer, and Cheng]{jiang2020atomic}
J.~Jiang, A.~Mishkin, K.~Prasai, M.~Yazback, R.~Zhang, R.~Bassiri, M.~Fejer and
  H.~Cheng, \emph{https://dcc.ligo.org/LIGO-P1900371}\relax
\mciteBstWouldAddEndPuncttrue
\mciteSetBstMidEndSepPunct{\mcitedefaultmidpunct}
{\mcitedefaultendpunct}{\mcitedefaultseppunct}\relax
\EndOfBibitem
\end{mcitethebibliography}
\bibliographystyle{rsc} %the RSC's .bst file

\end{document}